\documentstyle[11pt]{article}
\catcode`\@=11
\begin{document} \openup6pt

\title{  EMERGENT UNIVERSE WITH EXOTIC MATTER}

\author{ S. Mukherjee\thanks{Email: sailom@iucaa.ernet.in}     
 and B.C. Paul\thanks{Email: bcpaul@iucaa.ernet.in }
  \\
Physics Department, North Bengal University \\
Dist : Darjeeling, PIN : 734 430, India. \\
N. K. Dadhich\thanks{Email : nkd@iucaa.ernet.in} \\
Inter-University Centre for Astronomy and Astrophysics, Pune, India 411007  \\
S. D. Maharaj\thanks{E-mail: maharaj@ukzn.ac.za} \\
 Astrophysics and Cosmology Research Unit  \\
School of  Mathematical Sciences, University of KwaZulu-Natal \\
Durban 4041, South Africa \\
and \\
A. Beesham\thanks{E-mail: abeesham@pan.uzulu.ac.za} \\
 Department of Mathematical Sciences, Zululand University \\
Kwadlangezwa, South Africa }

\date{}

\maketitle
\vspace{0.1cm}

\begin{abstract}

A general framework for an emergent universe scenario has been given which makes use of an equation of  state. The general features of the model have also been studied and possible primordial  composition of the universe have been suggested. 

\end{abstract}

\section{Introduction : }

The possibilities of an emergent universe have been studied recently in a number of papers [1-3] in which one looks for a universe which is ever-existing and  large enough  so that the space-time may be treated as classical entities. There is no time-like singularity. In these models, the universe in the infinite past is in an almost static state but it eventually evolves into an inflationary stage. These ideas are in conformity with the Lemaitre-Eddington concepts forwarded in the early days of modern cosmology, although the details and the context are different now. An emergent universe model, if developed in a consistent way is capable of solving the well known conceptual problems of the Big-Bang model. A model of an ever-existing universe, which eventually entered at some stage into the standard Big Bang epoch and consistent with features precisely known to us, will be of considerable interest. The purpose of this letter is to examine the possibilities of such a scenario.

We mention here three models of the emergent universe, which are relevant here.

1. A closed universe containing radiation and a cosmological constant, given by Harrison [4], which  contains the scale factor
\begin{equation}
a(t) = a_{i} \left[ 1 + exp \left( \frac{\sqrt{2} t}{a_{i}} \right) \right]^{1/2}.
\end{equation}
As $t \rightarrow - \infty$, the model goes over asymptotically to an Einstein static universe. Although ever inflating, at any time $t_{o} >> a_{i}$, the expansion is given by a finite number of e-folds,
\begin{equation}
N_{o} = \ln \left( \frac{a_{o}}{a_{i}} \right) \sim  \frac{t_{o}}{\sqrt{2}}{a_{i}}.
\end{equation}

2. The second example has been studied by Ellis and Maartens [1]  and  Ellis {\it et al.} [2]. They considered a closed universe containing a minimally coupled scalar field  $\phi$, which has a self-interaction given by a special potential function $V (\phi)$. This potential looks similar to what one obtains in a   $R 
+ \alpha R^{2}$ theory [5] after the conventional  conformal transformation   and identifying  the scalar field  $\phi$     as  $ \phi = - \sqrt{3} \; \ln (1 + 2 \alpha R) $ with $\alpha$ negative. Although the solution is not obtained analytically, the model exhibits features expected in an emergent universe.

3. A third example has been provided by Mukherjee {\it et al.} [3], where it was shown  that  the Starobinsky Model has a solution which can describe an emergent universe. Here, one considers the semiclassical Einstein equation,
\begin{equation}
R_{\mu \nu} - \frac{1}{2} g_{\mu \nu} R = - \; 8 \pi G < T_{\mu \nu} >
\end{equation}
where    $< T_{\mu \nu} > $ is the vacuum expectation value of the energy momentum tensor of the fields. Assuming only free, massless and conformally invariant fields, and a Robertson-Walker metric, one obtains, for a spatially flat universe, the following equation :
\begin{equation}
H^{2} \left( \frac{1}{ K_{3}} - H^{2} \right) =  -  \frac{ 6 K_{1}}{K_{3}} \left( 2 H \ddot{H} + 6 H^{2} \dot{H} - \dot{H}^{2} \right),
\end{equation}
where the constant $K_{3}$ is determined by the species and number of fields and $K_{1}$ is a constant which may be chosen freely. It has been shown that the equation permits a solution which describes an emergent universe, with a scale factor
\begin{equation}
a(t) = a_{i} \left( \beta  + e^{\alpha t} \right)^{2/3}.
\end{equation}
where $ \alpha = \frac{3}{2} \sqrt{ \frac{1}{K_{3}}} $ and $ K_{1} = 
- \frac{2}{27} K_{3}$, $\beta$ is an integration constant. The general features of the model have been given in  [3].

These examples indicate that solutions, describing an emergent universe, occur in different contexts. It will, therefore, be interesting to see if solutions describing an emergent universe can be classified and studied in a general way. A simple approach will be to look for the equation of state (EOS) which lead to such solutions. In the next section we obtain such an EOS and study the general features of the relevant solutions, without referring to the actual source of the energy density. In section 3, we shall consider various combinations of radiation and matter, normal or exotic. We discuss our results in the last section.

\section{ The Equation of State (EOS) for emergent universe }

The cosmological models usually consider linear equation of state (EOS),  viz. $p = \omega \rho$, where $\omega $ is a constant, depending on the nature of the constituents. A notable exception is the case of a scalar field. For a homogeneous scalar field $\phi$ interacting with a potential $V( \phi) $, $$ \omega = \frac{\frac{1}{2} \dot{\phi}^{2} - V(\phi) }{\frac{1}{2} \dot{\phi}^{2} + V(\phi) } $$
where $\omega$ may vary between - 1 and +1. 
Tachyonic condensates provide another case of varying $\omega$, e.g.
\begin{equation}
\omega =  - \left(1 - \dot{\phi}^{2} \right)
\end{equation}
where $\omega$ is always negative.

Recent astronomical data when interpreted in the context of the Big Bang model have provided some interesting information about the composition of the universe. The total energy density has three components, while  Big Bang nucleosynthesis data suggest that baryonic matter can account for only about  4 $\%$  of the total energy density,  the cold dark matter (CDM) is about 23 $\%$ and  the third  part,  called dark energy, constitutes the remaining 73 $\%$.
The CDM has an almost dustlike EOS and it is considered to be responsible for clustering on galactic or supergalactic scales. The dark energy on the other hand provides a negative pressure which may explain the  recent acceleration in the  expansion of the universe in the context of a closed universe. The behaviour of dark energy  is very similar to that of a cosmological constant. 

In looking for a model of the emergent universe, we assume the following features for the universe :

1) The universe is isotropic and homogeneous at large scales. \\
2) It is spatially flat, as indicated by BOOMERANG and WMAP results. \\
3) It is ever existing. There is no singularity. \\
4) The universe is always large enough so that a classical description of space-time is adequate. \\
5) The matter or in general, the source of gravity has to be described by quantum field theory. \\
6) The universe may contain exotic matter so that energy conditions may be violated. \\
7) The universe is accelerating as suggested by Supernova observations. \\

The presence of  exotic components indicate that we need to revise our concepts about the primordial composition of the universe and the EOS  need some generalisation. In the following, we consider the EOS
\begin{equation}
p (\rho) = A \rho - B \rho^{\frac{1}{2} }
\end{equation}
where $A$ and $B$ are constant. The energy density $ \rho$ may have different components, each satisfying its own equation of state.

The Einstein equations for a flat universe in RW metric are given by
\begin{equation}
\rho = 3 \frac{ \dot{a}^{2}}{a^{2}},
\end{equation}
\begin{equation}
p  =  - 2  \frac{ \ddot{a}}{a} - \frac{ \dot{a}^{2}}{a^{2}}.
\end{equation}
Making use of eqs (7) - (9) , we get the equation, with $ \dot{a} >0$,
\begin{equation}
 2  \frac{ \ddot{a}}{a} + ( 3 A + 1)  \frac{ \dot{a}^{2}}{a^{2}}  - \sqrt{3} B \frac{ \dot{a}}{a} = 0
\end{equation}
which can be integrated once to give
\begin{equation}
\dot{a} a^{ \frac{  3 ( A + 1)}{2}} = K e^{- \frac{ \sqrt{3}}{2} B t}
\end{equation}
leading to the solution
\begin{equation}
a ( t ) = \left(  \frac{ 3 K (A + 1)}{2} \left( \sigma +  \frac{2}{\sqrt{3 }B } 
e^{\frac{\sqrt{3}}{2}  B t } \right) \right)^{\frac{2}{3 (A + 1)}}
\end{equation}
where $K$ and $\sigma$ are two constants of integration. We note the following :

1) If $ B < 0$, the solution has a singularity and it is not of interest to us here.

2) If $B > 0$, the solution describes an emergent universe if $A > -1$. The solution in this case can be written as,
\begin{equation}
a ( t ) = a_{i} \left(  \beta + e^{\alpha t} \right)^{\omega}
\end{equation}
 where $a_{i}$ and $\beta$ are constants, $\alpha = \frac{\sqrt{3}}{2} B$, and $ \omega =  \frac{2}{3 (A + 1)}$. 
The Hubble parameter and its derivatives are given by
\begin{equation}
H = \frac{\omega \alpha e^{\alpha t}}{\beta + e^{\alpha t}}, \; \dot{H} = 
\frac{\beta \omega \alpha^{2} e^{\alpha t}}{(\beta + e^{\alpha t})^{2}}, \; \ddot{H} = \frac{\beta \omega \alpha^{3} e^{\alpha t} (\beta -  e^{\alpha t})}{(\beta + e^{\alpha t})^{3}}
\end{equation} 
Here $H$ and $\dot{H}$ are both positive, but $\ddot{H}$ changes sign at  $t = \frac{1}{\alpha} \; \ln \beta$.  Thus $H$, $\dot{H}$ and $\ddot{H}$ both tend to zero as $t \rightarrow - \infty$. On the otherhand as $t \rightarrow  \infty$ the solution gives asymptotically a de Sitter universe. $\beta$ can be determined if the time when  $\dot{H}$ is a maximum, can be fixed from observational data.

3) The solution with Starobinsky model, obtained by Mukherjee {\it et al.} [3] , appears to be  a special solution with $A =  0$, $B > 0$. However, Starobinsky model is based on a semi-classical Einstein equation and in the initial stage, we have no matter and the vacuum energy of the fields act as the source of gravitation.  It is indeed a solution of a different equation and in that sense it  does not belong to the class of solutions considered here.

\section{ Composition of the emergent universe }

To study the possible composition of the emergent universe, we first study the dependence of the energy density on the scale factor. Consider the energy conservation equation,
\begin{equation}
\dot{\rho} + 3 (\rho + p) \frac{ \ddot{a}}{a}  = 0.
\end{equation}
Making use of the EOS, equation (7), and integrating we obtain the relation
\begin{equation}
\rho ( a )  =  \frac{ 1}{(A + 1 )^{2}}     \left[ B +  \frac{ K}{ a^{3/2 (A+1)}} \right]^{2}.
\end{equation}
where $K$ is an integration constnt.
Since $a$ is a monotonically increasing function of $t$ in the model,  one may use $a$ to study  the evolution of the universe also. 

It may be pointed out that a minimally coupled scalar field cannot give rise to the emergent universe of the type we are considering here (spatially flat, expanding, accelerating and singularity free). To see this  we note that the conservation eq. (15) leads to the field equation
\begin{equation}
\ddot{\phi} + 3 H \dot{\phi} + \frac{dV(\phi)}{d\phi} = 0
\end{equation}
where $V(\phi)$ is the self-interaction potential of the field $\phi$. However, 
$\rho + p = \dot{\phi}^{2}$ must be positive.  We, therefore,
require $\dot{\rho} < 0$. But  we have $ \dot{\rho} = 6 H \dot{H}$ which is always positive in this model. Inclusion of a cosmological constant will not change the conclusion.  Note that the solutions of  [1] and [2] were obtained for a closed universe. Cosmological  studies have made extensive use  of scalar fields and, therefore,  the emergent scenario has naturally been left out of  consideration. However, if the recent observations of spatial features and the presence of both dark matter and dark energy are confirmed, one should look for alternative sources and the possibility of an emergent universe containing exotic matter cannot be ruled out.

The equation (16) provides us the information about the components of energy density that lead to an emergent universe.  We can rewrite the equation as
\begin{equation}
\rho = \frac{ B^{2}}{(A + 1 )^{2}} +  \frac{ 2 K B}{  (A+1)^{2}}  
\frac{ 1}{ a^{3/2 (A+1)}} + \frac{  K^{2}}{  (A+1)^{2}}  
\frac{ 1}{ a^{3(A+1)}} \\
= \rho_{1} + \rho_{2} + \rho_{3}.
\end{equation}
The pressure $p$, given by equation (7), can also be expressed similarly,
\begin{equation}
p = - \frac{ B^{2}}{(A + 1 )^{2}} +  \frac{ B K (A - 1)}{  (A+1)^{2}}  
\frac{ 1}{ a^{3/2 (A+1)}} + \frac{  A K^{2}}{  (A+1)^{2}}  
\frac{ 1}{ a^{3(A+1)}} \\
= p_{1} + p_{2} + p_{3}.
\end{equation}
It is now easy to identify the components :

a) The first term behaves like a cosmological constant and may account for  the dark energy.

b) The second and  third terms depend on the choice of the parameters. Thus if $A = \frac{1}{3}$, we have $p_{2} = - \frac{1}{3}\rho_{2}$, which describes cosmic strings and   $p_{3} =  \frac{1}{3}\rho_{3}$, which describes radiation and ultra relativistic particles. Thus, it seems that an emergent universe could have evolved out of a mixture of cosmic strings and radiation, along with a cosmological constant.

Other possibilities also exist as shown in Table 1.  For a given  cosmological constant $\Lambda  = \left (\frac{B}{A + 1} \right)^{2}$, we have different possible compositions with different kinds of matter and radiation. Topological defects cosmic string,  and domain walls have already been studied in the context of structure formation in the early universe and it is interesting to note that emergent universe can accommodate these exotic energy sources.

If $A$ is given a value very close to 1, we have $p_{2} \sim 0$, corresponding to a dust like matter and $p_{3} = \rho_{3}$, describing a stiff matter. The stiff mater component falls rapidly with the scale factor, $\rho_{3} \sim \frac{1}{a^{6}}$, and  unless particle interactions change the relevant EOS, one may observe only the dark energy and the dust like matter part. Note that we have four parameters in this theory $A, B$, $a_{i}$ and  $\beta$ . As the present observational data indicate that the total energy density has three components, three of these parameters are  determined. The  measurement of the scale factor at any time or 
when the universe is quasi-static with $a_{s} \sim a_{i} \beta^{\omega}$,   will  determine the fourt parameter.

The composition may change, as in the standard Big Bang cosmology, due to  non-gravitational interactions. 

\begin{table}[htbp]
\centerline{\footnotesize Table 1.}
\begin{tabular}{l c c c c c c} \\
\hline \\
A   &  $\frac{\rho_{2}}{ \Lambda }$ in unit $\frac{ K}{B}$ & $\omega_{2}$ & 
$ \frac{\rho_{3}}{ \Lambda}  $ in unit $(\frac{ K}{B})^{2}$  & $\omega_{3}$ & Composition \\
\\
\hline \\
$ \frac{1}{3}$ &\phantom0$\frac{9}{8a^{2}}$ & $- \frac{1}{3}$ & $\frac{9}{8a^{4}}$ & $ \frac{1}{3}$& dark energy, \\
{} & & & &  &  cosmic string and radiation \\
\hline \\
$ - \frac{1}{3}$ &\phantom0$\frac{9}{2a}$  &   $- \frac{2}{3}$ & $\frac{9}{4a^{2}}$ & $-  \frac{1}{3}$& dark energy \\
{} & & & &  &  domain wall and cosmic string  \\
\hline \\
$1$ &\phantom0$\frac{1}{2a^{3}}$ & 0 & $\frac{1}{4a^{6}}$  & 1 & dark energy, \\
{} & & & &  & dust and stiff matter  \\
\hline \\
$ 0$ &\phantom0$\frac{2}{8a^{3/2}}$  &   $- \frac{1}{2}$ & $\frac{1}{a^{3}}$  & 0& dark energy,\\
{} & & & &  & exotic matter and dust   \\
\hline \\
\end{tabular}
\caption{ \it  Composition of universal matter for various   values of A }
\end{table}

\vspace{0.5 cm.}

\section{DISCUSSION}

We have shown in this letter that emergent universe scenarios are not isolated solutions and they may occur for different combinations of radiation and matter. The recipe for an emergent universe for a given cosmological constant (Dark energy), $\Lambda = \left (\frac{B}{A + 1} \right)^{2}$, has been given in Table 1. The exotic matter mentioned in the last line of the table ($A = 0$), which has an EOS, $p =- \frac{1}{2} \rho$, is not yet known. It may be an unstable energy source and may have decayed into other particles or radiation. The possibility of cosmic string or domain walls  playing a role in the evolution of the universe has been studied previously  in detail [6] and it is interesting to note that these topological defects suitably combined also lead to an emergent universe.  Cosmic string in particular could serve as seeds for galaxy formation and larger scale structure formation. This should also be observable through their gravitational lensing and studies of anisotropy microwave background radiation and the background gravitational waves etc. However, the  scenario of the phase transition of the relevant scalar field leading to these topological defects remains to be worked out in this model. It will be interesting to try to develop an evolutionary scenario of the emergent universe and this is presently under our consideration.

\vspace{0.5 cm.}

{\large {\it  Acknowledgments :}}

SM and BCP would like to thank the University of Zululand and the University of 
KwaZulu-Natal, South Africa for hospitality during their visit when a part of the work 
was done. They would also like to thank IUCAA, Pune  and IUCAA Reference Centre, North Bengal 
University for providing facilities. 

\pagebreak

\end{document}